\documentstyle[epsf,aps,psfig]{revtex}
\renewcommand{\thefootnote}{\fnsymbol{footnote}}
\begin{document}
\setcounter{footnote}{1}
\begin{center}
{\Large\bf Debye screening and Meissner effect in a three-flavor 
color superconductor}
\\[1cm]
Dirk H.\ Rischke 
\\ ~~ \\
{\it RIKEN-BNL Research Center} \\
{\it Brookhaven National Laboratory, Upton, New York 11973, U.S.A.} \\
{\it email: rischke@bnl.gov}
\\ ~~ \\ ~~ \\
\end{center}
\begin{abstract} 
I compute the gluon self-energy in a color superconductor with
three flavors of massless quarks, where condensation of Cooper pairs 
breaks the color and flavor $SU(3)_c \times U(3)_V \times U(3)_A$ symmetry
of QCD
to the diagonal subgroup $SU(3)_{c+V}$. At zero temperature,
all eight electric gluons obtain a Debye screening mass, and all
eight magnetic gluons a Meissner mass. The Debye as well as the
Meissner masses are found to be equal for the different gluon colors.
These masses determine the coefficients of the kinetic terms
in the effective theory for the low-energy degrees of freedom.
Their values agree with those obtained by Son and Stephanov.
\end{abstract}
\renewcommand{\thefootnote}{\arabic{footnote}}
\setcounter{footnote}{0}

\section{Introduction and Conclusions}

The presence of attractive interactions in a degenerate fermionic
system destabilizes the Fermi surface and leads to the formation
of Cooper pairs --- the system becomes a superconductor \cite{BCS}. 
When increasing the quark density in cold quark matter,
asymptotic freedom implies that
single-gluon exchange becomes the dominant interaction between
quarks. Single-gluon exchange is attractive in the color-antitriplet channel, 
and therefore leads to color superconductivity in cold, dense quark matter 
\cite{bailinlove}.

Considerable activity has been recently generated 
\cite{ARW,cfl,rajetal,ABRmeissner,RSSV,SW,cont,SWQCD,hsu,hsuQCD,rdpdhrlett1,rdpdhrscalar,rdpdhr2,rdpdhrparity,rdpdhrproc,rob,dhr2f,carter,hong,hongetal,OhioQCD,shovkovy,son,sonstephanov,rockef,hongzahed,RWZ,RSWZ,casalbuonigatto,bedaque,BBS,lang,matsu,gorbar,zarembo,sannino,andi,chandra}
by the observation that
the zero-temperature color-superconducting gap $\phi_0$
can be as large as 100 MeV \cite{ARW,RSSV}.
Gaps of this magnitude may have important consequences 
for the physics of nuclear collisions.
The critical temperature for the onset
of color superconductivity, $T_c$, is (to leading order in the
strong coupling constant $g$) related to $\phi_0$ in the same
way as in BCS theory, $T_c \simeq 0.57\, \phi_0$ \cite{rdpdhr2}.
Thus, for $\phi_0 \sim 100$ MeV, it cannot be excluded
that color-superconducting quark matter could be created in nuclear
collisions in the GSI--SIS or BNL--AGS energy range. 

In order to compute the color-superconducting gap, one commonly 
solves a gap equation \cite{BCS}. 
For theories with point-like four-fermion interactions,
$\phi_0 \sim \mu \, \exp(-c_{\rm 4F}/G^2)$, where $\mu$ is the
quark chemical potential, $c_{\rm 4F}$ is a constant, and $G^2$ is
the four-fermion coupling strength 
\cite{ARW,cfl,rajetal,RSSV,SW,hsu,carter,lang}. On the other
hand, in QCD, $\phi_0 \sim \mu \, \exp(-c_{\rm QCD}/g)$, where
$c_{\rm QCD}$ is another constant
\cite{SWQCD,hsuQCD,rdpdhr2,hong,OhioQCD,shovkovy,son,rockef,RWZ,RSWZ}.
In weak coupling, there are then three different energy scales,
$\phi_0 \ll m_g \ll \mu$, where $m_g$ is the gluon mass.
At $T=0$ and for $N_f$ flavors of massless quarks \cite{LeBellac},
\begin{equation}
m_g^2 \equiv \frac{N_f}{3}\, \frac{g^2 \mu^2}{2\pi^2}\,\, .
\end{equation}

The value of $c_{\rm QCD}$ depends on the form of the gluon propagator 
in the cold, dense quark medium. If one takes the gluon propagator
in the standard ``hard dense loop`` (HDL) approximation \cite{LeBellac},
one obtains $c_{\rm QCD} = 3\, \pi^2/\sqrt{2}$ \cite{son}.
In this approximation, the quarks inside the HDL's are assumed to be
in the normal, and not the color-superconducting phase.
Consequently, an important question that has to be addressed is how 
color superconductivity influences the propagation of
gluons and whether this could change $c_{\rm QCD}$. 
In a recent work \cite{dhr2f}, I have derived a general expression 
for the gluon self-energy in a two-flavor color superconductor, and
explicitly computed the self-energy in the static limit, $p_0 = 0$, for gluon
momenta $p \equiv |{\bf p}| \rightarrow 0$, and for $p \gg \phi_0$.

In a two-flavor color superconductor, condensation of Cooper pairs in a
channel of total spin $J=0$ breaks $SU(3)_c$ to $SU(2)_c$. 
Then, the three gluons 
corresponding to the generators of the unbroken $SU(2)_c$ subgroup
are expected to remain massless, while the other five should
attain a mass through the Anderson--Higgs mechanism. 
An explicit computation of the gluon self-energy to one-loop order in
perturbation theory confirms this qualitative expectation, but 
quantitatively reveals some surprising details \cite{dhr2f}. At $T=0$,
the three gluons of the unbroken $SU(2)_c$ attain no Meissner mass, 
but also no Debye mass. 
This means that static, homogeneous color-electric fields of the unbroken
$SU(2)_c$ subgroup are not screened. Furthermore, the Debye and Meissner
masses of the remaining five gluons are not identical: 
four gluons have a Debye mass $\sqrt{3/2}\,m_g$ and a Meissner
mass $m_g/\sqrt{2}$, while the last one has a Debye mass $\sqrt{3}\, m_g$,
like in the non-superconducting phase, but
a Meissner mass $m_g/\sqrt{3}$.

On the other hand, in a three-flavor color superconductor 
the color and flavor $SU(3)_c \times U(3)_V \times U(3)_A$
symmetry is broken to the diagonal subgroup $SU(3)_{c+V}$.
This locks color and flavor rotations \cite{cfl}.
From the 18 Goldstone bosons resulting from symmetry breaking,
eight get ``eaten'' by the gluons, which consequently become massive.
The purpose of this paper is to complement the results of \cite{dhr2f}
for the two-flavor case
with the computation of the Debye and Meissner masses of these gluons
in the three-flavor case.

It is worthwhile mentioning that here, as well as in \cite{dhr2f},
the terms ``Debye mass'' and ``Meissner mass'' refer exclusively
to the screening of {\em color\/}-electric and {\em color\/}-magnetic
fields. The ``ordinary'' (electro-)magnetic Meissner effect was studied
in \cite{ABRmeissner,gorbar}. Similar to the mixing of
weak and electromagnetic gauge bosons in the standard model, 
the electromagnetic field mixes with the eighth gluon to form a
modified photon, under which the color-superconducting
condensate is electrically neutral. The mass of the
modified eighth gluon becomes slightly larger than that of the other
seven. However, the mixing angle as well as this increase in mass
is determined by the ratio of electromagnetic 
and strong coupling constants and consequently quite small. 
Therefore, effects from electromagnetism will be neglected 
throughout the following.

There is another reason why it is important to know the
values for the Debye and Meissner mass. 
The relevant degrees of freedom
in the color-flavor locked phase at energy scales much smaller
than the gap, $\phi_0$, are the remaining 
10 Goldstone bosons resulting from the breaking of
color and flavor symmetries. Apart from an additional Goldstone boson
arising from breaking $U(1)_V$, these bosons are analogous to the 
pseudoscalar mesons which result from
chiral symmetry breaking in the QCD vacuum \cite{cont,rdpdhrproc}.
Consequently, the dynamics of these 10 Goldstone bosons
is described by an effective theory which
resembles the Lagrangian of the nonlinear sigma model,
describing the dynamics of the chiral fields in the QCD 
vacuum \cite{sonstephanov,casalbuonigatto}.
To lowest order, this Lagrangian contains only kinetic terms,
\begin{equation} \label{efftheory}
{\cal L}_{{\rm nl}\Sigma}^{\rm kin} = 
\frac{f_\pi^2}{4} \, {\rm Tr} \left( \partial_0 
\Sigma^\dagger \partial_0 \Sigma - v_\pi^2 \bbox{\nabla} \Sigma^\dagger
\cdot \bbox{\nabla} \Sigma \right) 
+ 12\, f_{\eta'}^2 \, \left[ \left( \partial_0 \theta \right)^2
- v_{\eta'}^2 \left( \bbox{\nabla} \theta \right)^2 \right]
+ 12\, f_{H}^2 \, \left[ \left( \partial_0 \varphi \right)^2
- v_{H}^2 \left(\bbox{\nabla} \varphi \right)^2 \right]\,\, .
\end{equation}
Here, $\Sigma \equiv \exp ( i\, \lambda^a/f_\pi)$, where 
$\pi^a,\, a=1,\ldots,8,$ are the fields
corresponding to the meson octet in the QCD vacuum. These are the
pions, the kaons, and the $\eta$ meson, which are the Goldstone
bosons resulting from spontaneous breaking of the $SU(3)_A$ symmetry. 
$\lambda^a$ are the Gell-Mann matrices.
$\theta$ is the field corresponding
to the meson singlet in the QCD vacuum, {\it i.e.},
the $\eta'$ meson, which is the
Goldstone boson resulting from breaking $U(1)_A$
spontaneously. (In the QCD vacuum, $U(1)_A$ is also broken
explicitly by instantons. At the quark densities
relevant for the effective theory in the color-flavor locked phase, however,
instantons play no longer any significant role \cite{rdpdhrlett1}.) Finally,
$\varphi$ is the Goldstone mode resulting from breaking $U(1)_V$.
This field has no analogon in the QCD vacuum. 

The coefficients of the kinetic terms are determined by the 
``decay constants'' $f_\pi$, $f_{\eta'}$, and $f_H$.
The presence of a medium breaks Lorentz invariance, and 
the coefficients of the time-like and space-like terms may differ, {\it i.e.},
the velocities $v_\pi^2$, $v_{\eta'}^2$, and $v_H^2$
are not necessarily equal to one. These velocities
enter the dispersion relation of the Goldstone bosons as
$\epsilon_i^2(k) = v_i^2 \,k^2 + m_i^2$, $i = \pi,\,  \eta', \, H$.

The decay constants as well as the velocities can be computed by
matching the effective theory to the underlying microscopic theory.
A convenient way to do this was proposed by Son and Stephanov 
\cite{sonstephanov}. First, they observed that, by minimally
gauging the model (\ref{efftheory}), for instance the pseudoscalar
decay constant $f_\pi$ is related to the Debye 
mass of the gluons, $m_D$, via
\begin{equation}
f_\pi \equiv m_D / g \,\, ,
\end{equation}
and the velocity of the pseudoscalar mesons
is determined from the ratio of Debye and Meissner masses,
\begin{equation}
v_\pi \equiv m_M / m_D \,\, .
\end{equation}
The problem is thus reduced to computing these masses in the
underlying theory.
In this case, one has two choices. First, one may use QCD
in the color-superconducting ground state. 
This theory has quasiparticle as well
as quasi-antiparticle excitations, and is valid at all energy scales.
Second, one may start from the effective theory proposed by Hong \cite{hong}.
This theory contains only quasiparticle excitations around the 
Fermi surface, quasi-antiparticle
excitations have already been integrated out. It is valid at energy scales
which are much smaller than the chemical potential, $\mu$, but which can
be larger than the gap, $\phi_0$.

Son and Stephanov \cite{sonstephanov} used the latter theory to compute 
the Debye and Meissner masses. Their results are
\begin{equation} \label{SS}
m_D^2 = m_g^2\, \frac{21 - 8\, \ln 2}{18} 
 \;\;\;\; , \;\;\;\;
m_M^2 = m_g^2\, \frac{21 - 8\, \ln 2}{54} 
\,\, .
\end{equation}
Consequently, the velocity of the pseudoscalar mesons is
$v_\pi = 1/\sqrt{3}$. Note that the (square of the) Debye mass 
in the three-flavor superconductor, $m_D^2 \simeq 0.859\;m_g^2$,
is reduced by a factor 3.5 as compared to its 
value in a normal, cold medium, $m_D^2 = 3\, m_g^2$.

The result (\ref{SS}) is not undisputed throughout the literature. 
For instance, Rho, Shuryak, Wirzba, and Zahed \cite{RSWZ}
computed the Debye and Meissner masses from the gluon self-energy
in the full theory, including quasi-antiparticle excitations.
They obtained [see Eqs.\ (A.72) and (A.75) of \cite{RSWZ}],
\begin{equation}
m_D^2 = \frac{1}{2}\, m_g^2
\;\;\;\; , \;\;\;\;
m_M^2 = \frac{5}{6} \, m_g^2 \,\, .
\end{equation}
This is quite surprising, as it implies that the velocity of the
pseudoscalar mesons is superluminous, $v_\pi \equiv m_M/m_D = \sqrt{5/3}$.
Other results that can be found in the literature
are those of Zarembo \cite{zarembo}, which agree 
with Son and Stephanov's calculation. 
Beane, Bedaque, and Savage \cite{BBS} agree with
Son and Stephanov on the Debye and Meissner masses up to a factor of 2.

The second goal of this paper is to resolve this ambiguity in
the literature. The Debye and Meissner masses will
be computed in the full theory, {\it i.e.},
QCD in the color-superconducting ground state. The framework for
such a computation was already established in \cite{dhr2f}. 
As shown in the following section, the results are found to
agree with those of Son and Stephanov, Eq.\ (\ref{SS}).

The Debye and Meissner masses are not only important for
the nonlinear version (\ref{efftheory})
of the effective low-energy theory in a three-flavor color superconductor.
As outlined in \cite{dhr2f} they 
also determine the coefficients of
the kinetic terms in the {\em linear\/} version of the effective theory,
\begin{equation} \label{efftheory2}
{\cal L}_{{\rm l}\Sigma}^{\rm kin} 
= \alpha_{\rm e} \sum_{h=r,\ell} {\rm Tr} \left[
\left( D_0 \Phi_h \right)^\dagger \, D^0 \Phi_h \right] 
+ \alpha_{\rm m} \sum_{h=r,\ell} 
{\rm Tr} \left[ \left( D_i \Phi_h \right)^\dagger \, 
D^i \Phi_h \right]  \,\, .
\end{equation}
Since there is no reason why right- and left-handed terms should
differ in normalization, I assumed $\alpha_{{\rm e}\, r}
\equiv \alpha_{{\rm e}\, \ell} \equiv \alpha_{\rm e}$, and similarly for
the coefficient of the space-like terms, $\alpha_{\rm m}$.
For color-flavor locking, the order parameter is a $3 \times 3$ matrix
with expectation value $\langle \Phi_h \rangle = {\rm diag}(\phi_{0\, h}, 
\, \phi_{0\, h},\, \phi_{0\, h})$ \cite{rdpdhrlett1}.
Consequently, $g^2 \alpha_{\rm e}( \phi_{0\, r}^2 + 
 \phi_{0\, \ell}^2) \equiv m_D^2$,
$g^2 \alpha_{\rm m} ( \phi_{0\, r}^2 + 
\phi_{0\, \ell}^2) \equiv m_M^2$. 
Due to explicit symmetry breaking by nonzero quark masses (and, at
less than asymptotically high densities, instantons), the true ground
state of the color-flavor locked phase corresponds to the $J^P=0^+$ channel
where $\phi_{0\,r} \equiv - \phi_{0\,\ell} \equiv \phi_0$.
Then, $\alpha_{\rm e} \equiv m_D^2/(2g^2 \phi_0^2)$, 
$\alpha_{\rm m}^2 \equiv m_M^2/(2g^2 \phi_0^2)$.

I use natural units, $\hbar=c=k_B=1$, and
work in Euclidean space-time ${\bf R}^4 \equiv V/T$, where $V$ is the volume
and $T$ the temperature of the system. Nevertheless, I find it convenient to
retain the Minkowski notation for 4-vectors, with a metric tensor
$g^{\mu \nu} = {\rm diag}(+,-,-,-)$. For instance, the space-time
coordinate vector is $x^\mu \equiv (t,{\bf x})$, $t \equiv -i\tau$, where
$\tau$ is Euclidean time. 4-momenta are denoted as
$K^\mu \equiv (k_0,{\bf k})$, $k_0 \equiv -i \omega_n$, where
$\omega_n$ is the Matsubara frequency, $\omega_n \equiv 2n \pi T$
for bosons and $\omega_n \equiv (2n+1) \pi T$ for fermions, $n=0,\pm 1,\pm 2,
\ldots$. The absolute value of the 3-momentum ${\bf k}$ is denoted as
$k \equiv |{\bf k}|$, and its direction as $\hat{\bf k} \equiv {\bf k}/k$.

\section{Explicit computation of Debye and Meissner masses}

A convenient starting point to compute the gluon self-energy
in the color-flavor locked phase is Eq.\ (68) of \cite{dhr2f},
\begin{eqnarray}
\Pi^{\mu \nu}_{ab} (P) & = & \frac{1}{2} \,g^2 \, \frac{T}{V} 
\sum_K {\rm Tr}_{s,c,f} \left[ \Gamma^\mu_a \, G^+ (K) \, \Gamma^\nu_b 
\, G^+(K-P) + \bar{\Gamma}^\mu_a \, G^- (K) \, \bar{\Gamma}^\nu_b 
\, G^-(K-P) \right. \nonumber \\
&   & \left. \hspace*{2.55cm}
+ \Gamma^\mu_a \, \Xi^- (K) \, \bar{\Gamma}^\nu_b 
\, \Xi^+(K-P) + \bar{\Gamma}^\mu_a \, \Xi^+ (K) \, \Gamma^\nu_b 
\, \Xi^-(K-P)  \right] \,\,. \label{Pi}
\end{eqnarray}
Here, the trace is over color, flavor, and spinor space.
The vertices are $\Gamma^\mu_a \equiv \gamma^\mu T_a$ and
$\bar{\Gamma}^\mu_a \equiv - \gamma^\mu T_a^T$.
$G^{\pm}$ and $\Xi^\pm$ are the diagonal and off-diagonal elements
of the Nambu--Gor'kov propagator for quasiparticle excitations,
\begin{equation}
G^\pm  \equiv  \left( G_0^\pm - \Sigma^\pm \right)^{-1}
\;\;\;\; , \;\;\;\;
\Xi^\pm  \equiv  - G_0^\mp \, \Phi^\pm \, G^\pm \,\, .
\end{equation}
$G_0^\pm(K) \equiv \left( \gamma \cdot K \pm \gamma_0 \mu \right)^{-1}$
is the propagator for massless,
non-interacting quarks (charge-conjugate quarks), 
and $\Sigma^\pm \equiv \Phi^\mp \, G_0^\mp \, \Phi^\pm $ 
is the quark self-energy generated by exchanging particles
or charge-conjugate particles with the condensate. 
In mean-field approximation, the condensate $\Phi^+$ is computed
from the gap equation discussed in \cite{rdpdhr2},
and $\Phi^-$ can be obtained from
\begin{equation} \label{phi-phi+}
\Phi^-(K) \equiv \gamma_0 \left[ \Phi^+(K) \right]^\dagger \gamma_0\,\, .
\end{equation}

\begin{figure}
\begin{center}
\epsfxsize=16cm
\epsfysize=4cm
\leavevmode
\hbox{ \epsffile{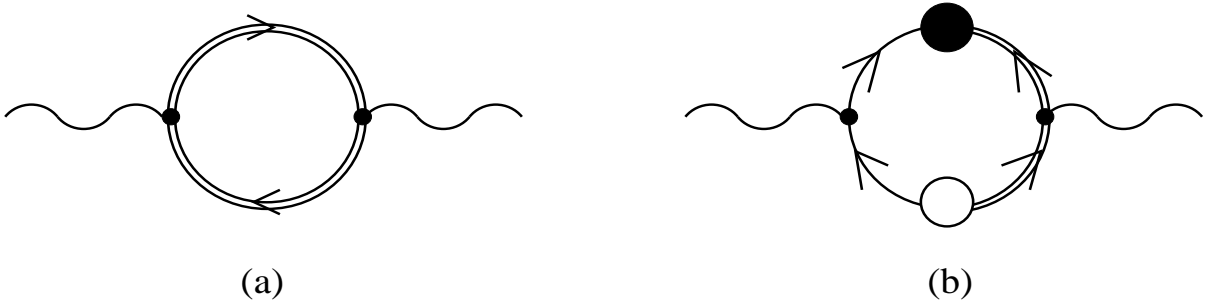}}
\end{center}
\caption{The contributions from (a) the diagonal and (b) the off-diagonal
components of the Nambu--Gor'kov propagator to the gluon self-energy.
Double full lines stand for the full quasiparticle propagator $G^\pm$,
single full lines for the free propagator $G_0^\pm$. The full blob
represents $\Phi^-$, the empty blob $\Phi^+$. Vertices are represented
by small blobs.}
\label{fig1}
\end{figure}

In Eq.\ (\ref{Pi}), the first line corresponds to the diagram in
Fig.\ \ref{fig1}(a), where only the diagonal
components of the Nambu--Gor'kov propagator appear, while the
second line corresponds to the diagram in Fig.\ \ref{fig1}(b),
formed from the off-diagonal components.
Note that, at small temperatures $T \sim \phi_0$, and
in weak coupling, $\phi_0 \ll \mu$, the fermion
loops of Fig.\ \ref{fig1} constitute the dominant contribution to
the gluon self-energy, since
contributions from gluon (or ghost) loops are relatively
suppressed by a factor $T^2/\mu^2 \sim \phi_0^2/\mu^2$ \cite{dhr2f}.

In the two-flavor case, the trace over color and flavor in Eq.\ 
(\ref{Pi}) could be
performed independently \cite{dhr2f}. In the three-flavor case,
due to color-flavor locking this is no longer possible.
The traces over color and flavor must be performed simultaneously.
The most elegant way to do this is to utilize the 
color-flavor space projectors
introduced by Shovkovy and Wijewardhana, Eq.\ (7) of \cite{shovkovy},
\begin{mathletters} \label{colorflavorproj}
\begin{eqnarray}
{{\cal C}^{(1)}}^{ij}_{rs} & \equiv & \frac{1}{3}\, \delta^i_r \, \delta^j_s
 \,\, , \\
{{\cal C}^{(2)}}^{ij}_{rs} & \equiv & \frac{1}{2}\left( \delta_{rs} \,
 \delta^{ij} - \delta^j_r \, \delta^i_s \right) \,\, , \\
{{\cal C}^{(3)}}^{ij}_{rs} & \equiv & \frac{1}{2}\left( \delta_{rs} \,
 \delta^{ij} + \delta^j_r \, \delta^i_s \right) - 
 \frac{1}{3}\, \delta^i_r \, \delta^j_s \,\, .
\end{eqnarray}
\end{mathletters}
(To avoid proliferation of the symbol ${\cal P}$, I denote them here
as ${\cal C}$.) All projectors are symmetric under simultaneous
exchange of color, $i,j$, and flavor, $r,s$, indices. 
Note that ${\cal C}^{(1)}$ is the singlet projector
${\bf P_1}$ introduced by Zarembo, Eq.\ (3.11) in \cite{zarembo}.
Furthermore, ${\cal C}^{(2)} + {\cal C}^{(3)} \equiv
{\bf 1} - {\bf P_1} \equiv {\bf P_8}$ is Zarembo's octet projector,
Eq.\ (3.12) in \cite{zarembo}.

With the projectors (\ref{colorflavorproj}), the gap matrices
$\Phi^\pm$ can be written as
\begin{equation} \label{gapcfproj}
\Phi^\pm \equiv \sum_{n=1}^3 {\cal C}^{(n)} \, \Phi_n^\pm \,\, .
\end{equation}
Here,
\begin{mathletters} \label{gaps}
\begin{eqnarray}
\Phi^\pm_1 & \equiv & 2 \left( \Phi_{\bf \bar{3}}^\pm + 2\, 
\Phi_{\bf 6}^\pm \right) \,\, , \\
\Phi^\pm_2 & \equiv &  \Phi_{\bf \bar{3}}^\pm -
\Phi_{\bf 6}^\pm \,\, , \\
\Phi^\pm_3 & \equiv & - \Phi^\pm_2 \,\, , \label{gaps3}
\end{eqnarray}
\end{mathletters}
are gap matrices in spinor space,
\begin{equation}
\Phi^+_n(K) \equiv \sum_{h=r,\ell} \sum_{e =\pm} \phi^e_{n,h}(K)\,
{\cal P}_h\, \Lambda^e_{\bf k} \;\;\;\; , \;\;\;\;
\Phi^-_n(K) \equiv \sum_{h=r,\ell} \sum_{e =\pm} 
\left[\phi^e_{n,h}(K)\right]^*\, {\cal P}_{-h}\, \Lambda^{-e}_{\bf k}
\,\, ,
\end{equation}
where ${\cal P}_{r,\ell} \equiv (1 \pm \gamma_5)/2$ are
chirality projectors, $-h = \ell$ when $h=r$ and $-h = r$ when
$h = \ell$, $\Lambda^\pm_{\bf k} 
\equiv (1 \pm \gamma_0 \bbox{\gamma} \cdot \hat{\bf k})/2$
are energy projectors, and $\phi^e_{n,h}(K)$ are
simple functions of 4-momentum $K^\mu$. 

In Eqs.\ (\ref{gaps}), $\Phi_{\bf \bar{3}}^\pm$ is the gap matrix in the
antitriplet channel and $ \Phi_{\bf 6}^\pm$ is the gap matrix in
the sextet channel, 
\begin{equation}
{\Phi^\pm}^{ij}_{rs} \equiv \Phi_{\bf \bar{3}}^\pm \left(
\delta^i_r \, \delta^j_s - \delta^i_s \, \delta^j_r \right)
+ \Phi_{\bf 6}^\pm \left(
\delta^i_r \, \delta^j_s + \delta^i_s \, \delta^j_r \right) \,\, .
\end{equation}
Why condensation in the (repulsive) sextet channel is possible 
in the color-flavor locked phase was explained in \cite{rdpdhrproc}.
Antitriplet and sextet gaps are related to the gap functions
$\kappa_1$ and $\kappa_2$ of \cite{cfl} by
$\phi_{{\bf \bar{3}}\, \ell}^+ = - \phi_{{\bf \bar{3}}\, r}^+ 
\equiv (\kappa_1-\kappa_2)/2$,
$\phi_{{\bf 6}\, \ell}^+ = - \phi_{{\bf 6}\, r}^+
\equiv (\kappa_1+\kappa_2)/2$. 
For future purpose, it will be convenient to define a singlet and an octet
gap matrix according to
\begin{equation} \label{singletoctetgaps}
\Phi_{\bf 1}^\pm \equiv \Phi_1^\pm \;\;\;\; , \;\;\;\;
\Phi_{\bf 8}^\pm \equiv \Phi_2^\pm \equiv - \Phi_3^\pm \,\, .
\end{equation}

The quasiparticle propagators take the form
\begin{equation} \label{prop}
G^\pm (K) \equiv \sum_{n=1}^3 {\cal C}^{(n)}\, G_n^\pm(K) \,\, ,
\end{equation}
where
\begin{equation} \label{quasiparticleprop}
G_n^\pm (K) = \sum_{h = r, \ell} \sum_{e = \pm} {\cal P}_{\pm h}\,
\Lambda^{\pm e}_{\bf k}\, 
\frac{1}{k_0^2 - [\epsilon^e_{\bf k}(\phi_{n,h}^e)]^2} \,
\left[ G_0^\mp(K) \right]^{-1} \,\, .
\end{equation}
The quasiparticle energies are
\begin{equation} \label{quasiparticleenergy}
\epsilon_{\bf k}^e ( \phi_{n,h}^e ) \equiv \sqrt{(\mu - ek)^2 + 
| \phi_{n,h}^e |^2} \,\, ,
\end{equation}
where $\phi_{n,h}^e$ is the gap function for pairing of quarks ($e = +1$)
or antiquarks ($e= -1$) with chirality $h$.

The right-hand side of (\ref{quasiparticleprop})
does not depend on the sign of the gap functions, {\it i.e.},
the difference in sign between $\phi^e_{2,h}$ and $\phi^e_{3,h}$, 
Eq.\ (\ref{gaps3}), is irrelevant, $G_2^\pm \equiv G_3^\pm$.
Then, Eq.\ (\ref{prop}) has the alternative representation
\begin{equation}
G^\pm(K) \equiv {\bf P_1} \, G_{\bf 1}^\pm(K) + {\bf P_8}\,
G_{\bf 8}^\pm(K) \,\, ,
\end{equation}
where ${\bf P_{1,8}}$ are the singlet and octet projectors introduced
by Zarembo \cite{zarembo}, see above, and, following
Eq.\ (\ref{singletoctetgaps}), $G_{\bf 1}^\pm \equiv
G_1^\pm$, $G_{\bf 8}^\pm \equiv G_2^\pm = G_3^\pm$.

The off-diagonal components of the quasiparticle propagators are
similarly computed as
\begin{equation} \label{offprop}
\Xi^\pm (K) \equiv \sum_{n=1}^3 {\cal C}^{(n)}\, \Xi_n^\pm(K) \,\, ,
\end{equation}
where
\begin{equation}
\Xi_n^+(K) = - \sum_{h = r,\ell} \sum_{e=\pm}
\frac{\phi^e_{n,h}(K)}{k_0^2 - [ \epsilon_{\bf k}^e
(\phi^e_{n,h})]^2} \, {\cal P}_{-h} \,
\Lambda^{-e}_{\bf k} \;\;\; , \;\;\;\;
\Xi_n^-(K) = - \sum_{h = r,\ell} \sum_{e=\pm}
\frac{\left[\phi^e_{n,h}(K)\right]^*}{k_0^2 - [ \epsilon_{\bf k}^e
(\phi^e_{n,h})]^2} \, {\cal P}_h \, \Lambda^e_{\bf k} \,\, .
\end{equation}
Since $\phi^e_{2,h} = - \phi^e_{3,h}$,
there is no simple representation in terms
of singlet and octet projectors for $\Xi^\pm$.
Nevertheless, in line with (\ref{singletoctetgaps})
let us define for future purpose
$\Xi_{\bf 1}^\pm \equiv \Xi_1^\pm$, $\Xi_{\bf 8}^\pm \equiv
\Xi_2^\pm \equiv - \Xi_3^\pm$.

Inserting Eqs.\ (\ref{prop}) and (\ref{offprop}) into Eq.\ (\ref{Pi}),
one straightforwardly performs the trace over color and flavor space to 
obtain
\begin{mathletters} \label{Pimunu}
\begin{eqnarray}
\Pi^{\mu \nu}_{ab} (P)&  = & \delta_{ab}\, \Pi^{\mu \nu}(P) \,\, ,  \\
\Pi^{\mu \nu}(P) & = & \frac{g^2}{12} \, \frac{T}{V} 
\sum_K {\rm Tr}_{s} \left[ \gamma^\mu \, G_{\bf 1}^+ (K) \, \gamma^\nu 
\, G_{\bf 8}^+(K-P) + \gamma^\mu \, G_{\bf 8}^+ (K) \, \gamma^\nu 
\, G_{\bf 1}^+(K-P)\frac{}{}  \right. \nonumber \\
&   & \left. \hspace*{1.8cm}
+ \;\gamma^\mu \, G_{\bf 1}^- (K) \, \gamma^\nu 
\, G_{\bf 8}^-(K-P) + \gamma^\mu \, G_{\bf 8}^- (K) \, \gamma^\nu 
\, G_{\bf 1}^-(K-P) \frac{}{} \right. \nonumber \\
&   & \left. \hspace*{1.6cm}
+\; 7\, \gamma^\mu \, G_{\bf 8}^+ (K) \, \gamma^\nu 
\, G_{\bf 8}^+(K-P) + 7\, \gamma^\mu \, G_{\bf 8}^- (K) \, \gamma^\nu 
\, G_{\bf 8}^-(K-P) \frac{}{} \right. \nonumber \\
&   & \left. \hspace*{1.8cm}
+ \;\gamma^\mu \, \Xi_{\bf 1}^- (K) \, \gamma^\nu
\, \Xi_{\bf 8}^+(K-P) + \gamma^\mu \, \Xi_{\bf 8}^- (K) \, \gamma^\nu 
\, \Xi_{\bf 1}^+(K-P) \frac{}{}  \right. \nonumber \\
&   & \left. \hspace*{1.8cm}
+ \;\gamma^\mu \, \Xi_{\bf 1}^+ (K) \, \gamma^\nu
\, \Xi_{\bf 8}^-(K-P) + \gamma^\mu \, \Xi_{\bf 8}^+ (K) \, \gamma^\nu 
\, \Xi_{\bf 1}^-(K-P) \frac{}{}  \right. \nonumber \\
&   & \left. \hspace*{1.6cm}
+ \;2\, \gamma^\mu \, \Xi_{\bf 8}^- (K) \, \gamma^\nu
\, \Xi_{\bf 8}^+(K-P) + 2\, \gamma^\mu \, \Xi_{\bf 8}^+ (K) \, \gamma^\nu 
\, \Xi_{\bf 8}^-(K-P)\frac{}{}  \right] \,\,. \label{Pimunu2}
\end{eqnarray}
\end{mathletters}
From (\ref{Pimunu}) one learns two things. First, unlike the
two-flavor case \cite{dhr2f}, the gluon self-energy is diagonal
in the adjoint colors $a,b$. Second, there are no terms where both
quasiparticle propagators involve singlet gaps. The reason is that such terms
are proportional to ${\rm Tr}_c T_a\, {\rm Tr}_c T_b \equiv 0$.

The evaluation of the spin traces proceeds in complete analogy to
the two-flavor case \cite{dhr2f}. Assuming $\phi^e_{n,r} = - \phi^e_{n,\ell}
\equiv \phi^e_n \in {\bf R}$, the result is [cf.\ Eq.\ (96a) of \cite{dhr2f}]
\begin{eqnarray}
\lefteqn{ \Pi^{\mu \nu} (P) =  - \frac{g^2}{12}\, 
 \int \frac{d^3{\bf k}}{(2 \pi)^3} \sum_{e_1,e_2 = \pm} 
 \left( \frac{}{}\!
{\cal T}_+^{\mu \nu}({\bf k}_1,{\bf k}_2) \right. } \label{Pimunu3} \\
& \times &  \left[ \left( \frac{\hat{n}_1 \, (1-n_2)}{p_0 + \hat{\epsilon}_1
+ \epsilon_2} - \frac{(1-\hat{n}_1)\, n_2}{p_0 - \hat{\epsilon}_1
- \epsilon_2} \right) (1-\hat{N}_1 - N_2) 
+ \left( \frac{(1-\hat{n}_1)\, (1-n_2)}{p_0 - \hat{\epsilon}_1
+ \epsilon_2} - \frac{\hat{n}_1\, n_2}{p_0 + \hat{\epsilon}_1
- \epsilon_2} \right) (\hat{N}_1 - N_2) \right. \nonumber \\
& +   &  \;\; \left( \frac{n_1 \, (1-\hat{n}_2)}{p_0 + \epsilon_1
+ \hat{\epsilon}_2} - \frac{(1-n_1)\, \hat{n}_2}{p_0 - \epsilon_1
- \hat{\epsilon}_2} \right) (1-N_1 - \hat{N}_2) 
+ \left( \frac{(1-n_1)\, (1-\hat{n}_2)}{p_0 - \epsilon_1
+ \hat{\epsilon}_2} - \frac{n_1\, \hat{n}_2}{p_0 + \epsilon_1
- \hat{\epsilon}_2} \right) (N_1 - \hat{N}_2)
\nonumber \\
& +   & 7 \left.  \!
\left( \frac{n_1 \, (1-n_2)}{p_0 + \epsilon_1
+ \epsilon_2} - \frac{(1-n_1)\, n_2}{p_0 - \epsilon_1
- \epsilon_2} \right) (1-N_1 - N_2) 
+ 7\, \left( \frac{(1-n_1)\, (1-n_2)}{p_0 - \epsilon_1
+ \epsilon_2} - \frac{n_1\, n_2}{p_0 + \epsilon_1
- \epsilon_2} \right) (N_1 - N_2)
 \right]  \nonumber \\
&   & \hspace*{4.3cm} + \;\;
 {\cal T}_-^{\mu \nu}({\bf k}_1,{\bf k}_2) \nonumber \\
& \times &  \left[ \left( \frac{(1-\hat{n}_1)\,n_2)}{p_0 + \hat{\epsilon}_1
+ \epsilon_2} - \frac{\hat{n}_1\,(1- n_2)}{p_0 - \hat{\epsilon}_1
- \epsilon_2} \right) (1-\hat{N}_1 - N_2) 
+ \left( \frac{\hat{n}_1\, n_2}{p_0 - \hat{\epsilon}_1
+ \epsilon_2} - \frac{(1-\hat{n}_1)\,(1- n_2)}{p_0 + \hat{\epsilon}_1
- \epsilon_2} \right) (\hat{N}_1 - N_2) \right. \nonumber \\
& +   &  \;\; \left( \frac{(1-n_1) \, \hat{n}_2}{p_0 + \epsilon_1
+ \hat{\epsilon}_2} - \frac{n_1\, (1-\hat{n}_2)}{p_0 - \epsilon_1
- \hat{\epsilon}_2} \right) (1-N_1 - \hat{N}_2) 
+ \left( \frac{n_1\,\hat{n}_2}{p_0 - \epsilon_1
+ \hat{\epsilon}_2} - \frac{(1-n_1)\,(1- \hat{n}_2)}{p_0 + \epsilon_1
- \hat{\epsilon}_2} \right) (N_1 - \hat{N}_2)
\nonumber \\
& +   & 7 \left.  \!
\left( \frac{(1-n_1) \, n_2}{p_0 + \epsilon_1
+ \epsilon_2} - \frac{n_1\, (1-n_2)}{p_0 - \epsilon_1
- \epsilon_2} \right) (1-N_1 - N_2) 
+ 7\, \left( \frac{n_1\, n_2}{p_0 - \epsilon_1
+ \epsilon_2} - \frac{(1-n_1)\,(1- n_2)}{p_0 + \epsilon_1
- \epsilon_2} \right) (N_1 - N_2)
 \right]  \nonumber \\
&   & \hspace*{4.3cm} - \;\left[ {\cal U}^{\mu \nu}_+({\bf k}_1, {\bf k}_2) + 
{\cal U}^{\mu \nu}_- ({\bf k}_1, {\bf k}_2) \right] 
\nonumber \\
& \times &  \left\{ \frac{\hat{\phi}_1\, \phi_2}{4 \, \hat{\epsilon}_1\, 
\epsilon_2} \left[ \left( \frac{1}{p_0 + \hat{\epsilon}_1 + \epsilon_2} 
- \frac{1}{p_0 - \hat{\epsilon}_1 - \epsilon_2} \right) 
( 1- \hat{N}_1 - N_2 )
- \left( \frac{1}{p_0 - \hat{\epsilon}_1 + \epsilon_2} - 
\frac{1}{p_0 + \hat{\epsilon}_1 - \epsilon_2} \right) 
( \hat{N}_1 - N_2 ) \right] 
\right. \nonumber \\
& + & \;\;\, \frac{\phi_1\, \hat{\phi}_2}{4 \, \epsilon_1\, \hat{\epsilon}_2} 
\left[ \left( \frac{1}{p_0 + \epsilon_1 + \hat{\epsilon}_2} - \frac{1}{
p_0 - \epsilon_1 - \hat{\epsilon}_2} \right) ( 1- N_1 - \hat{N}_2 )
- \left( \frac{1}{p_0 - \epsilon_1 + \hat{\epsilon}_2} 
- \frac{1}{p_0 + \epsilon_1- \hat{\epsilon}_2} \right)
( N_1 - \hat{N}_2 ) \right] 
\nonumber \\
& + &  \left.\left.\! 2\,\frac{\phi_1\, \phi_2}{4 \, \epsilon_1\, \epsilon_2} 
\left[ \left( \frac{1}{p_0 + \epsilon_1 + \epsilon_2} - \frac{1}{
p_0 - \epsilon_1 - \epsilon_2} \right) ( 1- N_1 - N_2 )
- \left( \frac{1}{p_0 - \epsilon_1 + \epsilon_2} - \frac{1}{p_0 + \epsilon_1
- \epsilon_2} \right) ( N_1 - N_2 ) \right] 
\right\} \right) . \nonumber
\end{eqnarray}
Here, I denoted the octet and singlet gaps by
\begin{equation}
\phi_i \equiv \phi^{e_i}_{\bf 8} 
\;\;\;\; , \;\;\;\;
\hat{\phi}_i \equiv \phi^{e_i}_{\bf 1} \,\, .
\end{equation}
Correspondingly,
\begin{equation}
\epsilon_i \equiv \epsilon_{{\bf k}_i}^{e_i}(\phi_i)
\;\;\;\; , \;\;\;\;
\hat{\epsilon}_i \equiv \epsilon_{{\bf k}_i}^{e_i}(\hat{\phi}_i)
\end{equation}
are the excitation energies for quasiparticles with octet and
singlet gaps, ${\bf k}_1 \equiv {\bf k}$,
${\bf k}_2 \equiv {\bf k} - {\bf p}$,
\begin{equation}
n_i \equiv n^{e_i}_{{\bf k}_i} \equiv \frac{\epsilon_i - \xi_i}{2\, \epsilon_i}
\;\;\;\; , \;\;\;\;
\hat{n}_i \equiv \hat{n}^{e_i}_{{\bf k}_i} \equiv 
\frac{\hat{\epsilon}_i - \xi_i}{2 \, \hat{\epsilon}_i}
\end{equation}
are the occupation numbers for quasiparticles with
octet and singlet gaps, $\xi_i \equiv e_i \, k_i - \mu$, and
\begin{equation}
N_i \equiv N^{e_i}_{{\bf k}_i} \equiv
\left[ \exp \left( \frac{\epsilon_i}{T} \right) +1 \right]^{-1} 
\;\;\;\; , \;\;\;\;
\hat{N}_i \equiv \hat{N}^{e_i}_{{\bf k}_i} \equiv
\left[ \exp \left( \frac{\hat{\epsilon}_i}{T} \right) +1 \right]^{-1} 
\end{equation}
are the corresponding thermal occupation numbers.
The spin traces are [see Eqs.\ (45) and (97) of \cite{dhr2f}]
\begin{mathletters} 
\begin{eqnarray}
{\cal T}_\pm^{00} & = & {\cal U}_\pm^{00} =
1 + e_1 e_2\, \hat{\bf k}_1 \cdot \hat{\bf k}_2
\,\, , \\
{\cal T}_\pm^{0i}  & =  & {\cal T}_\pm^{i0} = - {\cal U}_\pm^{0i}
= {\cal U}_\pm^{i0} =  \pm e_1 \, \hat{k}_1^i \pm e_2 \, \hat{k}_2^i 
\;\;\; , \;\;\;\; i = x,y,z \,\, , \\
{\cal T}_\pm^{ij} & = & - {\cal U}_\pm^{ij} = \delta^{ij} \, 
\left(1- e_1 e_2\,\hat{\bf k}_1 \cdot \hat{\bf k}_2\right) + e_1 e_2 \,
\left(\hat{k}_1^i \, \hat{k}_2^j + \hat{k}_1^j \, \hat{k}_2^i\right)
\;\;\; , \;\;\;\; i,j = x,y,z \,\, .
\end{eqnarray}
\end{mathletters}
In Eq.\ (\ref{Pimunu3}), the terms proportional to ${\cal T}_\pm^{\mu \nu}$
correspond to the diagram in Fig.\ \ref{fig1}(a), while those
proportional to ${\cal U}_\pm^{\mu \nu}$ correspond to that
in Fig.\ \ref{fig1}(b).

In order to compute the Debye and Meissner mass, it is sufficient
to consider the time-like, $\mu = \nu =0$, and space-like,
$\mu = i, \, \nu=j$, components of the self-energy, since
the Debye and Meissner masses are defined as
\begin{equation} \label{defineDM}
m_D^2 \equiv -\lim_{p \rightarrow 0} \Pi^{00}(0,p)\;\;\;\; ,\;\;\;\;
m_M^2 \equiv \lim_{p \rightarrow 0} \Pi^{ii}(0,p)\,\, .
\end{equation}
The sign in the first equation is due to the choice of metric.
The self-energy of electric gluons is
\begin{mathletters} \label{selfenergies}
\begin{eqnarray}
\Pi^{00}(P) & = & - \frac{g^2}{12} \int 
\frac{d^3{\bf k}}{(2 \pi)^3}  \sum_{e_1,e_2 = \pm} (1 + e_1 e_2\,
\hat{\bf k}_1 \cdot \hat{\bf k}_2) \nonumber \\
&  & \times \left[ \left( \frac{1}{p_0 + \hat{\epsilon}_1 + \epsilon_2}
- \frac{1}{p_0 - \hat{\epsilon}_1 - \epsilon_2} \right) 
( 1- \hat{N}_1 - N_2 ) \, 
\frac{ \hat{\epsilon}_1 \, \epsilon_2 - \xi_1 \, \xi_2 - \hat{\phi}_1\, 
\phi_2}{2 \, \hat{\epsilon}_1 \, \epsilon_2} \right. \nonumber \\
&   &  + \;\; \left( \frac{1}{p_0 + \epsilon_1 + \hat{\epsilon}_2}
- \frac{1}{p_0 - \epsilon_1 - \hat{\epsilon}_2} \right) 
( 1- N_1 - \hat{N}_2 ) \, 
\frac{ \epsilon_1 \, \hat{\epsilon}_2 - \xi_1 \, \xi_2 - \phi_1\, 
\hat{\phi}_2}{2 \, \epsilon_1 \, \hat{\epsilon}_2} \nonumber \\
&   & + \, 7 \left( \frac{1}{p_0 + \epsilon_1 + \epsilon_2}
- \frac{1}{p_0 - \epsilon_1 - \epsilon_2} \right) 
( 1- N_1 - N_2 ) \, 
\frac{ \epsilon_1 \, \epsilon_2 - \xi_1 \, \xi_2 - 2\,\phi_1\, 
\phi_2/7}{ 2 \, \epsilon_1 \, \epsilon_2} \nonumber \\
&  & + \;\; \left( \frac{1}{p_0 - \hat{\epsilon}_1 + \epsilon_2}
- \frac{1}{p_0 + \hat{\epsilon}_1 - \epsilon_2} \right) 
(\hat{N}_1 - N_2 ) \, 
\frac{ \hat{\epsilon}_1 \, \epsilon_2 + \xi_1\, \xi_2 + \hat{\phi}_1 \, 
\phi_2}{2 \, \hat{\epsilon}_1\,  \epsilon_2} \nonumber \\
&  & + \;\; \left( \frac{1}{p_0 - \epsilon_1 + \hat{\epsilon}_2}
- \frac{1}{p_0 + \epsilon_1 - \hat{\epsilon}_2} \right) 
(N_1 - \hat{N}_2) \, 
\frac{ \epsilon_1 \, \hat{\epsilon}_2 + \xi_1\, \xi_2 + \phi_1 \, 
\hat{\phi}_2}{2 \, \epsilon_1\,  \hat{\epsilon}_2} \nonumber \\
&  & + \left.\! 7 \left( \frac{1}{p_0 - \epsilon_1 + \epsilon_2}
- \frac{1}{p_0 + \epsilon_1 - \epsilon_2} \right) 
(N_1 - N_2 ) \, 
\frac{ \epsilon_1 \, \epsilon_2 + \xi_1\, \xi_2 + 2\, \phi_1 \, \phi_2/7}{
2 \, \epsilon_1\,  \epsilon_2} \right] \,\, .
\end{eqnarray}
On the other hand, the self-energy of magnetic gluons is
\begin{eqnarray}
\Pi^{ij}(P) & = & - \frac{g^2}{12} \int 
\frac{d^3{\bf k}}{(2 \pi)^3}  \sum_{e_1,e_2 = \pm} 
\left[ \delta^{ij} \left(1 - e_1 e_2\,\hat{\bf k}_1 \cdot \hat{\bf k}_2\right) 
+ e_1 e_2 \left( \hat{k}_1^i \, \hat{k}_2^j + \hat{k}_1^j \,
\hat{k}_2^i \right) \right] \nonumber \\
&  & \times \left[ \left( \frac{1}{p_0 + \hat{\epsilon}_1 + \epsilon_2}
- \frac{1}{p_0 - \hat{\epsilon}_1 - \epsilon_2} \right) 
( 1- \hat{N}_1 - N_2 ) \, 
\frac{ \hat{\epsilon}_1 \, \epsilon_2 - \xi_1 \, \xi_2 + \hat{\phi}_1\, 
\phi_2}{2 \, \hat{\epsilon}_1 \, \epsilon_2} \right. \nonumber \\
&   &  + \;\; \left( \frac{1}{p_0 + \epsilon_1 + \hat{\epsilon}_2}
- \frac{1}{p_0 - \epsilon_1 - \hat{\epsilon}_2} \right) 
( 1- N_1 - \hat{N}_2 ) \, 
\frac{ \epsilon_1 \, \hat{\epsilon}_2 - \xi_1 \, \xi_2 + \phi_1\, 
\hat{\phi}_2}{2 \, \epsilon_1 \, \hat{\epsilon}_2} \nonumber \\
&   & + \, 7 \left( \frac{1}{p_0 + \epsilon_1 + \epsilon_2}
- \frac{1}{p_0 - \epsilon_1 - \epsilon_2} \right) 
( 1- N_1 - N_2 ) \, 
\frac{ \epsilon_1 \, \epsilon_2 - \xi_1 \, \xi_2 + 2\,\phi_1\, 
\phi_2/7}{ 2 \, \epsilon_1 \, \epsilon_2} \nonumber \\
&  & + \;\; \left( \frac{1}{p_0 - \hat{\epsilon}_1 + \epsilon_2}
- \frac{1}{p_0 + \hat{\epsilon}_1 - \epsilon_2} \right) 
(\hat{N}_1 - N_2 ) \, 
\frac{ \hat{\epsilon}_1 \, \epsilon_2 + \xi_1\, \xi_2 - \hat{\phi}_1 \, 
\phi_2}{2 \, \hat{\epsilon}_1\,  \epsilon_2} \nonumber \\
&  & + \;\; \left( \frac{1}{p_0 - \epsilon_1 + \hat{\epsilon}_2}
- \frac{1}{p_0 + \epsilon_1 - \hat{\epsilon}_2} \right) 
(N_1 - \hat{N}_2) \, 
\frac{ \epsilon_1 \, \hat{\epsilon}_2 + \xi_1\, \xi_2 - \phi_1 \, 
\hat{\phi}_2}{2 \, \epsilon_1\,  \hat{\epsilon}_2} \nonumber \\
&  & + \left.\! 7 \left( \frac{1}{p_0 - \epsilon_1 + \epsilon_2}
- \frac{1}{p_0 + \epsilon_1 - \epsilon_2} \right) 
(N_1 - N_2 ) \, 
\frac{ \epsilon_1 \, \epsilon_2 + \xi_1\, \xi_2 - 2\, \phi_1 \, \phi_2/7}{
2 \, \epsilon_1\,  \epsilon_2} \right] \,\, .
\end{eqnarray}
\end{mathletters}
In Eq.\ (\ref{selfenergies}), terms proportional to 
$(\epsilon_1 \, \epsilon_2 \pm \xi_1\, \xi_2)/(2\, \epsilon_1\, \epsilon_2)$ 
(and similar terms involving the singlet gaps, $\hat{\phi}_i$)
arise from the diagram in Fig.\ \ref{fig1}(a),
while terms proportional to $\phi_1 \, \phi_2/(2\, \epsilon_1\, \epsilon_2)$ 
(and similar terms involving the singlet gaps, $\hat{\phi}_i$)
arise from that in Fig.\ \ref{fig1}(b).

To proceed I treat quasi-antiparticles as 
free antiparticles, as in \cite{dhr2f},
\begin{mathletters}
\begin{eqnarray}
& &\phi^-_{\bf 8} \simeq 0 \;\;\;\; , \;\;\;\;  \phi^-_{\bf 1} \simeq 0
\;\;\;\; , \;\;\;\;
\epsilon^-_{{\bf k}_i}(\phi^-_{\bf 8}) \simeq k_i + \mu \;\;\;\; , \;\;\;\;
\epsilon^-_{{\bf k}_i}(\phi^-_{\bf 1}) \simeq k_i + \mu\,\, , \\
& & n^-_{{\bf k}_i} \simeq 1 \;\;\;\; , \;\;\;\;  
\hat{n}^-_{{\bf k}_i} \simeq 1\;\;\;\; , \;\;\;\;
1-n^-_{{\bf k}_i} \simeq 0 \;\;\;\; , \;\;\;\; 
1-\hat{n}^-_{{\bf k}_i} \simeq 0 \;\;\;\; , \;\;\;\;
N^-_{{\bf k}_i} \simeq 0 \;\;\;\; , \;\;\;\; 
\hat{N}^-_{{\bf k}_i} \simeq 0\,\,.
\end{eqnarray}
\end{mathletters}
Setting $p_0 = 0$ and sending $p \rightarrow 0$,
there are then no antiparticle contributions to the
electric part of the gluon self-energy. Abbreviating
$\phi^+_{\bf 8} \equiv \phi$, $\phi^+_{\bf 1} \equiv \hat{\phi}$,
$\epsilon^+_{\bf k} \equiv \epsilon$, $\hat{\epsilon}^+_{\bf k} \equiv
\hat{\epsilon}$, $N^+_{\bf k} \equiv N$, $\hat{N}^+_{\bf k} \equiv \hat{N}$,
\begin{mathletters}
\begin{eqnarray}
\Pi^{00} (0) & \equiv & \Pi^{(a)}_{\rm e} (0)+ \Pi^{(b)}_{\rm e}(0) \,\, , \\
\Pi^{(a)}_{\rm e}(0) & \simeq & - \frac{g^2}{24 \pi^2} \int_0^\infty 
dk \, k^2 \, \left[ 8\, \frac{1-\hat{N}-N}{\hat{\epsilon} + \epsilon} 
 \, \frac{ \hat{\epsilon}\, \epsilon - \xi^2}{2\,\hat{\epsilon}\, \epsilon }
+  7\, (1 - 2\, N)\, \frac{ \phi^2}{\epsilon^3} \right. \nonumber \\
&   & \hspace*{2.6cm} 
- \left. 8\, \frac{N- \hat{N}}{\epsilon - \hat{\epsilon}} \,
\frac{\hat{\epsilon}\, \epsilon + \xi^2 }{2\,
\hat{\epsilon}\, \epsilon} - 28 \, \frac{{\rm d} N}{{\rm d} \epsilon} \,
\left( 1 - \frac{\phi^2}{2\,\epsilon^2} \right) \right] \,\, , \\
\Pi^{(b)}_{\rm e}(0) & \simeq & - \frac{g^2}{24 \pi^2} \int_0^\infty 
dk \, k^2 \, \left[ -\, 8\, \frac{1-\hat{N}-N}{\hat{\epsilon} + \epsilon} 
 \, \frac{ \hat{\phi}\, \phi }{2\,\hat{\epsilon}\, \epsilon }
- 2\, (1 - 2\, N)\, \frac{ \phi^2}{\epsilon^3} \right. \nonumber \\
&   & \hspace*{3cm} 
- \left. \! 8\, \frac{N- \hat{N}}{\epsilon - \hat{\epsilon}} \,
\frac{ \hat{\phi}\, \phi}{2\,\hat{\epsilon}\, \epsilon} 
- 4 \, \frac{{\rm d} N}{{\rm d} \epsilon} \,
\frac{\phi^2}{\epsilon^2} \right] \,\, .
\end{eqnarray}
In order to facilitate comparison with the results
of Son and Stephanov, and to elucidate the origin of the
various terms, I have separated the self-energy into the contributions
from the diagram in Fig.\ \ref{fig1}(a), $\Pi^{(a)}_{\rm e}$,
and that in Fig.\ \ref{fig1}(b), $\Pi^{(b)}_{\rm e}$.

For the self-energy of magnetic gluons one obtains with 
$\int d\Omega \, \hat{k}^i\, \hat{k}^j/(4 \pi) = \delta^{ij}/3$,
$n^+_{\bf k} \equiv n$, $\hat{n}^+_{\bf k}\equiv \hat{n}$,
\begin{eqnarray}
\Pi^{ij}(0) & \equiv & \delta^{ij}\, \left[\Pi^{(a1)}_{\rm m} (0) 
+ \Pi^{(a2)}_{\rm m} (0) + \Pi^{(b)}_{\rm m} (0)\right] \,\, ,\\
\Pi^{(a1)}_{\rm m} (0) & \simeq & - \frac{g^2}{72 \pi^2}
\int_0^\infty dk\, k^2 \, 
\left[ 8\, \frac{1-\hat{N}-N}{\hat{\epsilon} + \epsilon} 
 \, \frac{ \hat{\epsilon}\, \epsilon - \xi^2 }{2\,\hat{\epsilon}\, \epsilon }
+  7\, (1 - 2\, N)\, \frac{ \phi^2}{\epsilon^3} \right. \nonumber \\
&   & \hspace*{2.6cm} 
- \left. 8\, \frac{N- \hat{N}}{\epsilon - \hat{\epsilon}} \,
\frac{\hat{\epsilon}\, \epsilon + \xi^2 }{2\,\hat{\epsilon}\, \epsilon} 
- 28 \, \frac{{\rm d} N}{{\rm d} \epsilon} \,
\left( 1 - \frac{\phi^2}{2\, \epsilon^2} \right) \right]\,\, , \\
\Pi^{(a2)}_{\rm m} (0) & \simeq & - \frac{g^2}{72 \pi^2}
\int_0^\infty dk\, k^2 \, 
\left[ 16\, \frac{(1-\hat{N})(1-\hat{n})}{k + \mu + \hat{\epsilon}}
+ 128 \, \frac{(1-N)(1-n)}{k + \mu + \epsilon} \right. \nonumber \\
&   & \hspace*{2.6cm} \left.
+ \, 16 \, \frac{\hat{N}\, \hat{n}}{k + \mu - \hat{\epsilon}}
+ 128\, \frac{N\, n}{k + \mu - \epsilon} - 72\, \frac{1}{k} 
 \right] \,\, , \label{Pia2m} \\
\Pi^{(b)}_{\rm m}(0) & \simeq & - \frac{g^2}{72 \pi^2}
\int_0^\infty dk\, k^2 \, 
\left[ 8\, \frac{1-\hat{N}-N}{\hat{\epsilon} + \epsilon} 
 \, \frac{\hat{\phi}\, \phi }{2\,\hat{\epsilon}\, \epsilon }
+  2\, (1 - 2\, N)\, \frac{ \phi^2}{\epsilon^3} \right. \nonumber \\
&   & \hspace*{2.6cm} 
+ \left. 8\, \frac{N- \hat{N}}{\epsilon - \hat{\epsilon}} \,
\frac{\hat{\phi}\, \phi}{2\,\hat{\epsilon}\, \epsilon} 
+ 4 \, \frac{{\rm d} N}{{\rm d} \epsilon} \,
\frac{\phi^2}{\epsilon^2} \right] \,\, .
\end{eqnarray}
\end{mathletters}
Again, I have separated contributions from Fig.\ \ref{fig1}(a),
$\Pi^{(a1)}_{\rm m} + \Pi^{(a2)}_{\rm m}$, from those of Fig.\ \ref{fig1}(b),
$\Pi^{(b)}_{\rm m}$.
Obviously, $\Pi^{(a)}_{\rm e} \equiv 3\, \Pi^{(a1)}_{\rm m}$,
$\Pi^{(b)}_{\rm e} \equiv - 3\, \Pi^{(b)}_{\rm m}$.
There are two contributions from Fig.\ \ref{fig1}(a). The first,
$\Pi^{(a1)}_{\rm m}$, arises from quasiparticle-quasiparticle 
excitations, while the second, $\Pi^{(a2)}_{\rm m}$, originates
from quasiparticle-antiparticle excitations. The latter is
UV-divergent, and thus requires renormalization,
which is achieved by adding the last term in Eq.\ (\ref{Pia2m}).
As we shall see shortly, $\Pi^{(a2)}_{\rm m}$ gives rise 
to the ``bare'' Meissner mass
in Hong's effective theory \cite{hong,sonstephanov}, where
contributions involving antiparticles are integrated out.

Let us now consider the case of zero temperature, where
$N \equiv \hat{N} \equiv 0$. One may restrict the $k$ integration to
the region $0 \leq k \leq 2\mu$, the contribution from
$k \geq 2\mu$ can be shown to be negligible.
The various components of the self-energies are
\begin{mathletters}
\begin{eqnarray} 
\Pi^{(a)}_{\rm e}(0) & \equiv & 3\, \Pi^{(a1)}_{\rm m}(0) 
\simeq - \frac{g^2 \mu^2}{12 \pi^2}
\int_0^\mu {\rm d} \xi \, \left[ \frac{4}{\hat{\phi}^2 - \phi^2} \,
\left( \frac{\hat{\epsilon}^2+ \xi^2 }{\hat{\epsilon}}- 
\frac{\epsilon^2 + \xi^2 }{\epsilon} \right)
+ 7\, \frac{\phi^2}{\epsilon^3} \right] \,\, , \label{Pia} \\
\Pi^{(b)}_{\rm e}(0) & \equiv & -3\, \Pi^{(b)}_{\rm m}(0) 
\simeq - \frac{g^2 \mu^2}{12 \pi^2}
\int_0^\mu {\rm d} \xi \, \left[ -\, 
\frac{4\,\hat{\phi}\, \phi}{\hat{\phi}^2 - \phi^2} \,
\left( \frac{1}{\epsilon}- \frac{1}{\hat{\epsilon}} \right)
- 2\, \frac{\phi^2}{\epsilon^3} \right] \,\, ,\label{Pib} \\
\Pi^{(a2)}_{\rm m}(0) & \simeq &  \frac{g^2}{72 \pi^2}
\int_0^{2\mu} {\rm d} k \,k\, \left[ 8 \,
\frac{\mu \, (\hat{\epsilon}-k+ \mu) + \hat{\phi}^2}{\hat{\epsilon} \,
(\hat{\epsilon}+k+\mu)} + 64 \, 
\frac{\mu \, (\epsilon-k+ \mu) + \phi^2}{\epsilon\, (\epsilon+k+\mu)}
\right] \,\, . \label{Pia2mT0}
\end{eqnarray}
\end{mathletters}
The integral appearing in (\ref{Pia2mT0})
was already computed in \cite{dhr2f}, Eq.\ (122). The result is
\begin{equation}
\Pi^{(a2)}_{\rm m} (0) \simeq m_g^2\,\, .
\end{equation}
As advertised above, 
this is the ``bare'' Meissner mass appearing in Hong's effective
theory \cite{hong,sonstephanov}.

To obtain the expressions (\ref{Pia}) and (\ref{Pib}) 
for $\Pi^{(a)}_{\rm e}$ and $\Pi^{(b)}_{\rm e}$, I substituted 
$\xi \equiv k- \mu$ and exploited the
symmetry of the integrand around $\xi = 0$.
Furthermore, contributions $\sim \xi^2/\mu^2$ in
the integrands were neglected, because they give rise to terms of order
$\sim \phi^2/\mu^2$ relative to the leading terms.
Neglecting the momentum dependence of the gap function,
all remaining integrals are exactly solvable.
First note that, to leading order, $\int_0^\mu {\rm d} \xi \, 
\phi^2/\epsilon^3 \simeq 1$. This takes care
of the last term in Eqs.\ (\ref{Pia}) and (\ref{Pib}).
To compute the first, substitute
$y \equiv \ln [(\xi + \epsilon)/\phi]$ for $\xi$ in the first term
in parentheses, and $\hat{y} \equiv \ln [( \xi + \hat{\epsilon})/\hat{\phi}]$
in the second. 
Evaluating the $y$ integral, note that one must not approximate
$\ln [(\mu + \sqrt{\mu^2 + \phi^2})/\phi] \simeq \ln (2\mu/\phi)$
for the upper boundary of the $y$ integral, and similarly for
the $\hat{y}$ integral. The reason is that leading-order terms cancel
between these two integrals. The subleading terms conspire to
cancel the denominator $\hat{\phi}^2 - \phi^2$.
To leading order, one then obtains
\begin{equation}
\Pi^{(a)}_{\rm e}(0) \simeq - \frac{3}{2} \, m_g^2\;\;\;\; , 
\;\;\;\;
\Pi^{(b)}_{\rm e}(0) \simeq \frac{1}{3}\, m_g^2 \, \left(
1 + \frac{2 \,\hat{\phi}\,\phi}{\hat{\phi}^2 - \phi^2} \, 
\ln \frac{\hat{\phi}}{\phi} \right)\,\, .
\end{equation}
Neglecting the sextet gap, the singlet gap is twice the octet gap,
$\hat{\phi} = 2 \phi$, cf.\ Eq.\ (\ref{gaps}), and 
\begin{equation}
\Pi^{(b)}_{\rm e}(0) \simeq  \frac{1}{3} \, m_g^2 \left(
1 + \frac{4}{3}\, \ln 2 \right)\,\, .
\end{equation}
Using the definitions of the Debye and Meissner masses,
Eq.\ (\ref{defineDM}), this confirms 
the results of Son and Stephanov, Eq.\ (\ref{SS}), {\it q.e.d.}

\section*{Acknowledgements}

I thank R.\ Pisarski, T.\ Sch\"afer, and D.T.\ Son for 
discussions.
My thanks go to RIKEN, BNL and the U.S.\ Dept.\ of Energy for
providing the facilities essential for the completion of this work,
and to Columbia University's Nuclear Theory Group for
continuing access to their computing facilities.

\end{document}